\PassOptionsToPackage{table}{xcolor}
\documentclass[times]{GonnellaLab-StyleArXiv}

\usepackage{blindtext}
\usepackage[overload]{textcase}
\usepackage{amsmath,amssymb}
\usepackage{lmodern}
\usepackage{iftex}
\usepackage{enumitem}
\usepackage{pifont}
\usepackage{booktabs}
\usepackage{multicol}
\usepackage[T1]{fontenc}
\usepackage[utf8]{inputenc}
\usepackage{textcomp} 
\usepackage{color}
\usepackage{fancyvrb}
\usepackage{natbib}
\usepackage{xcolor}
\usepackage{multirow}
\usepackage{bigdelim}
\usepackage{makecell}
\usepackage{subcaption}

\IfFileExists{upquote.sty}{\usepackage{upquote}}{}
\IfFileExists{microtype.sty}{
  \usepackage[]{microtype}
  \UseMicrotypeSet[protrusion]{basicmath} 
}{}
\makeatletter
\@ifundefined{KOMAClassName}{
  \IfFileExists{parskip.sty}{%
    \usepackage{parskip}
  }{
    \setlength{\parindent}{0pt}
    \setlength{\parskip}{6pt plus 2pt minus 1pt}}
}{
  \KOMAoptions{parskip=half}}
\makeatother
\IfFileExists{xurl.sty}{\usepackage{xurl}}{} 
\IfFileExists{bookmark.sty}{\usepackage{bookmark}}{}
\hypersetup{
  hidelinks,
  pdfcreator={LaTeX via pandoc}}
\DefineVerbatimEnvironment{Highlighting}{Verbatim}{commandchars=\\\{\}}

\ifLuaTeX
  \usepackage{selnolig}  
\fi

\begin{document}

\leadauthor{Gonnella}

\title{ProSt: computing, storing and visualizing attributes of prokaryotic genomes}

\shorttitle{ProSt: prokaryotic genome attributes}

\author[1,2\space \Letter]{Giorgio Gonnella}

\affil[1]{Center for Bioinformatics (ZBH), Universität Hamburg, Bundesstrasse 43, 20146 Hamburg}
\affil[2]{Institute for Microbiology and Genetics, Georg-August-Universität Göttingen, Goldschmidtstr. 1, 37077 Göttingen}

\maketitle


\begin{abstract}
Prokaryotic organisms usually possess compact genomes, which are particularly suitable to complete sequencing with existing technologies,
which led to an escalating accumulation of available genome data.

In response to this ever-expanding repository of information, we introduce ProSt, a computational system designed for the batch computation,
storage, and interactive visualization of the values of attributes of prokaryotic genomes. The system allows for parallel attribute value batch computation, dynamically designed to incrementally integrate new attribute values as additional genomes become available.

ProSt is flexible permitting the definition of attributes by implementing attribute value computation plugins, supporting
several languages (Python, Nim, Rust and Bash). This allows the system to continually evolve in accordance with
changing research needs and developments. Additionally, our computation and storage systems
maintain comprehensive metadata, thereby enabling data provenance tracking for the computed attribute values.
\end {abstract}


\begin{corrauthor}
    giorgio.gonnella\at uni-goettingen.de
\end{corrauthor}


\begin{multicols}{2}
The number of fully sequenced prokaryotic genomes is exponentially increasing
thanks to the technological advancements in sequencing \citep{Zhang2020}.
These, together with the development of automated software tools for
sequence assembling and annotation, significantly reduced the cost and complexity
associated with prokaryotic genome sequencing.

In this context, comparative genomics applications become increasingly important
for understanding the biology of microorganisms \citep{Kobras2021}.
Recently, we introduced the EGC format for representing expectations about the contents
of prokaryotic genomes \citep{textformats, egc},
and a collection of such expectations derived from the
analysis of scientific literature \citep{collection}.

Genome attributes, as defined in \citet{unambiguous}, are variables
associated to one or multiple genome content units, i.e. sequence elements
or functional or structural elements of the genome, or products derived
from them (RNA, proteins).

Here, we present ProSt (short for ``prokaryotic statistics''),
a computational system for batch computing, storing, retrieving and visualizing
the values of genome attributes.

\section{Results}

\begin{figure*}
\begin{center}    
    \includegraphics[width=0.8\textwidth]{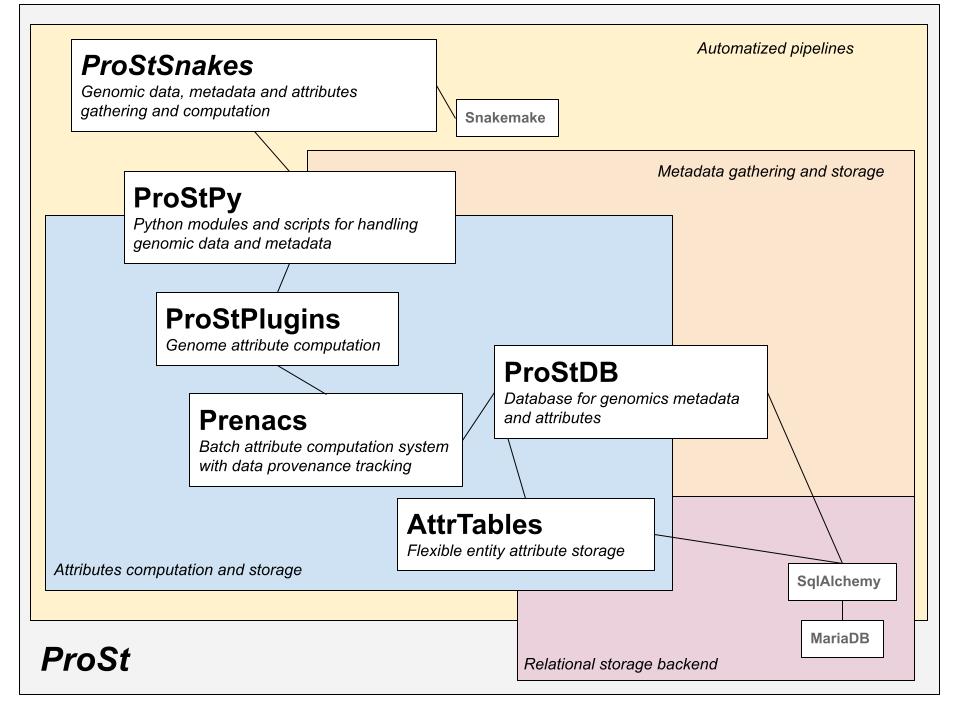}
    \caption{Architecture of the ProSt system}
    \label{fig:prostarch}
\end{center}
\end{figure*}

\subsection{Overview}

The ProSt software system is modular: it is implemented as a set of independent Python packages,
which are expected to be useful, independently of the ProSt system, in other applications.
All packages are open source, with their code available on GitHub,
and accompanied by user manuals and test suites.

The architecture of the system is summarized in Figure \ref{fig:prostarch}.
The components are summarized here and illustrated with more details in the following sections.
The main ProSt package is based on the MultiPlug, Prenacs and AttrTables packages.
Thereby, MultiPlug is a plugin system, which allows extending
existing Python programs, using different programming languages.
Prenacs is a batch computation system, which aims at being efficient and extensible,
enabling parallel computation and keeping track of data provenance.
AttrTables is a flexible database-based storage system for entity attributes. 

The highest level package of the ProSt system is represented by the ProSt package itself.
The package is organized in several components: a collection of Python scripts and modules
(ProStPy), Prenacs/MultiPlug computation plugins for genome attributes (ProStPlugins),
pipeline scripts for creating and keeping up-to-date a database of genome sequences, annotations
and associated attributes (ProStSnakes).

\subsection{The MultiPlug Python plugin system}

MultiPlug is a Python library which allow introducing in
any Python program a flexible plugin systems, supporting multiple programming languages.
Currently, plugins can be written in Python, Nim, Rust and Bash.
The library is generic, so that it can be applied not only to the system described in this
paper, but also in other contexts, whenever a plugin system is desirable.

An example usage of MultiPlug would be a command line interface script, which takes
the name of a module, with a predefined interface, as command line argument, and
execute the module code.

The library is available in Github at
\url{https://github.com/ggonnella/multiplug} and on the PyPi repository
at \url{https://pypi.org/project/multiplug/}. This text refers to version 1.2.

\subsubsection{Plugins implementation and execution}

In the plugin code, functions are defined using specialized libraries,
such as the \textit{nimpy} \citep{nimpy} and \textit{nimporter} \citep{nimporter} libraries for Nim and
the \textit{PyO3} \citep{PyO3} and \textit{maturin} libraries \citep{maturin} for Rust. Bash code must
follow conventions stated in the user manual, which allows for a wrapper code
to be automatically written by the library.

The code in the calling program is mostly independent of the plugin implementation language.
The only limitation to this are Bash plugins, which only support strings, lists
of strings, or lists of string lists as return type. Bash plugins can be disabled
if this limitation is not acceptable for an application.

Plugins are imported using the function \textit{multiplug.importer(plugin\_filename)},
which determines the implementation language form the plugin file extension.
An example is given in Figure \ref{fig:multiplug_run}.
The function optionally allows specifying the interface of the imported module,
as a list of constants and function which must or can be provided.

\begin{figure*}
\ \vspace{5mm}
\hrule
\ \vspace{0.5mm}
\begin{verbatim}
import multiplug

API = {"req_func": ["run"], "opt_func": ["before", "after"],
       "req_const": ["NAME"], "opt_const": ["NOTE"]}
plugin = multiplug.importer(args["<plugin>"], **API)

state = plugin.before() if plugin.before else None
for i in range(100):
  print(plugin.run(i, state))
if plugin.after:
  plugin.after(state)
\end{verbatim}
\ \vspace{-1mm}
\hrule
\ \vspace{-2mm}
\caption{Example usage of MultiPlug to import and run a plugin. The API of the plugin is described
in the API dictionary passed to the \texttt{importer} function. Thereby it is defined that the plugin
contains a \texttt{run} function, and may contain the functions \texttt{before} and \texttt{after},
as well as some metadata constants. The functions are then called by the subsequent code.}
\label{fig:multiplug_run}
\end{figure*}

\subsubsection{Defining module constants}

Python does not really have constants, since any value can be modified.
By convention, however, variables names written completely in upper case
are considered to be constants and their contents shall not be changed
by the user.

We desired to have a system for importing constants from the modules written
in different languages, so that we could defined special values. This
is important for our application, since it allows to store metadata
of the plugin, such as a plugin name, version number, and parameter information,
which are then stored in our database, for ensuring the proper
value provenance tracking.

In Python plugins, the constants are simply defined in the module code.
For the other languages, further implementation
effort and some conventions were necessary.

In Bash, constants are defined as variables in the script code.
Variables whose name does not start with an underscore are imported.
Special conventions are defined and described in the user manual
to import nested arrays of strings.

For Nim, we developed an own system, since the nimpy and nimporter libraries
do not allow to export Nim constants directly to Python.
Thus a workaround was necessary: thereby, constants are defined
as the return value of functions (which are exportable to Python).
We developed a Nim library \texttt{multiplug\_nim}, which automatically
defines these functions from normal Nim constants, using
a macro \texttt{exportpy\_consts()}.

A workaround is necessary also in Rust, since the PyO3 and maturin libraries
do not allow to export Rust module-level constants to Python.
In this case, constants are defined in a struct called \texttt{Constants}
and exported as Python class.

\subsubsection{Support of persistent state data}

In some cases it is useful to have an object, called \textit{state},
which persists after each call of the plugin functions. This is useful
when, for example, a library requires a complex data structure,
which does not need to be created for each single function call.

For this reason, we allow the plugins to define initialization and
finalization functions. Examples and explanations are given for each supported language.
For Python and Nim the implementation of this feature is straightforward:
the initialization function creates a state object, which is passed
using a special parameter name to other functions of the plugin.
For Rust plugins, the state must be initialized as a struct.
When using the state in other functions, the data ownership model of Rust
must be followed; thus the state must be borrowed and released accordingly.
In Bash plugins, persistent data can only be implemented by storing them to file.
Thus the initialization stores the state temporary file, which is then read
and eventually modified by other functions and deleted by the finalization function.

\begin{figure*}
\begin{small}
\hrule
\ \vspace{-3mm}
\begin{verbatim}\end{verbatim}
(a) Python \vspace{2mm}
\begin{verbatim}
import math

NAME = "sqrt.python"
NOTE = "uses the sqrt function of math"

def run(number):
  return math.sqrt(number)
\end{verbatim}
\ \vspace{0mm}
\hrule
\ \vspace{-3mm}
\begin{verbatim}\end{verbatim}
(b) Nim \vspace{2mm}
\begin{verbatim}
import math
import nimpy
import multiplug_nim/exportpy_consts

const
  NAME = "sqrt.nim"
  NOTE = "uses the sqrt function of std/math"
exportpy_consts(NAME, NOTE)

proc run(number: int): float {.exportpy.} =
  result = sqrt(number.float)
\end{verbatim}
\ \vspace{0mm}
\hrule
\ \vspace{-3mm}
\begin{verbatim}\end{verbatim}
(c) Bash \vspace{2mm}
\begin{verbatim}
NAME="sqrt.bash"
NOTE="uses the sqrt function of bc"

function run() { local number=$1;
  result=$(echo "scale=2;sqrt($number)" | bc)
  echo -n $result
}
\end{verbatim}
\ \vspace{0mm}
\hrule
\ \vspace{-3mm}
\begin{verbatim}\end{verbatim}
(d) Rust \vspace{2mm}
\begin{verbatim}
use std::num::sqrt;
use pyo3::prelude::*;
use pyo3::wrap_pyfunction;

#[pyclass] struct Constants {}
#[pymethods]
impl Constants {
  #[classattr] const NAME: &'static str = "sqrt.rust";
  #[classattr] const NOTE: &'static str = "uses sqrt of std::num";
}

#[pyfunction]
fn run(number: &f64) -> PyResult<f64> {
  let result: f64 = sqrt(number as f64);
  Ok(result)
}

#[pymodule]
fn sqrt_rust(_py: Python, m: &PyModule) -> PyResult<()> {
    m.add_class::<Constants>()?;
    m.add_function(wrap_pyfunction!(run, m)?)?;
    Ok(())
}
\end{verbatim}
\ \vspace{-1mm}
\hrule
\ \vspace{-2mm}
\end{small}

\caption{Example of MultiPlug plugins written in different languages: Python (a), Nim (b), Bash (c) and Rust (d). The \textit{run} function of each plugin takes a number as argument and
returns its square root.}
\label{fig:multiplug_examples}
\end{figure*}

\subsection{The Prenacs batch computation system}

The batch computation system for genome attributes used in ProSt is implemented as a separated Python package,
called Prenacs, acronym for ``PRovenance-tracking ENtities Attributes Computation System''.
Thereby, in the application described in this paper, entities are genomes and attributes are genome attributes.

However, the package can be applied to any set of objects, named here \textit{entities}.
The only requirement for entities is that they need to be uniquely identifiable
in the usage context of the program and that values can be computed for variables,
called \textit{attributes}, which describe some aspect of the entities.

For using Prenacs the following aspects
must be defined, which are described in the following sections: the description of
the attributes to be computed, the code for computing those attributes, and the input
set for the batch computation.

The Prenacs package is available in Github at
\url{https://github.com/ggonnella/prenacs} and on the PyPi repository
at \url{https://pypi.org/project/prenacs/}. This text refers to version 1.2.

\subsubsection{Defining and managing attributes}

Attribute are here defined as properties of the entity, for which a value can be determined.
The attribute can have any possible datatype and be of scalar or composite type.
The attribute system is conceived to be open and flexible, i.e.\ new attributes can be added at any moment.

Before running a computation, the attributes which are computed by the computation plugin must be defined
and added to the database, where the computation results are finally stored. Scripts are provided for this
task, as well as for editing or deleting existing attributes. These scripts accept as input an attributes metadata YAML file.
The file contains entries for each attribute, consisting of a free-text definition, a datatype description
(following an appropriate definition convention). The definition may also include the name of a computation group,
a set of attributes that are usually computer together: if provided, whenever the attributes are computed together,
the computation metadata is stored in common, reducing the storage overhead. Other optional metadata consists of
links to external references, e.g. ontology terms, descriptions of the measure units, and free-text remarks.

\subsubsection{Attribute computation plugins}

The code for the computation of attributes is provided to Prenacs as a plugin. Prenacs plugin are modules,
which compute the values of one or multiple attributes for a single entity of the input set.

The plugin implementations can be various, from thin wrappers over external programs or
scripts interfacing with external servers, to complex programs directly computing the attribute values.
Plugin implementations can be updated and improved. Thereby, thanks to a versioning system,
the provenance information of existing data, computed using the older
versions of the plugin is not lost.

Plugins are imported and executed using the MultiPlug library, and thus can be implemented in any of the
languages supported by that library: Python, Nim, Rust and Bash.
They all have a predefined interface (Table \ref{tab:plugins}), which includes a computation function and metadata constants, which describe the input, features and parameters
of the computation. Optionally functions can be defined, which initialize and finalize a state variable, persistent among
calls of the computation function. The computation function returns a tuple, including the results and computation logging messages.
A script is provided, which checks the conformance of plugins interface to the conventions.

\begin{table*}[b]
\centering
\rowcolors{2}{gray!20}{white}
\begin{tabular}{llp{92mm}}
\toprule
\texttt{ID} & required & unique identifier of the plugin \\
\texttt{VERSION} & required & version of the plugin \\
\texttt{INPUT} & required & type and format required for the input data \\
\texttt{OUTPUT} & required & list of names of the attributes computed by the plugin\\
\texttt{PARAMETERS} & poss.\ required & type, accepted values and
default value of each parameter (if any) \\
\texttt{METHOD} & optional & conceptually describes or gives a reference to the method used for the attributes computation \\
\texttt{IMPLEMENTATION} & optional & technical detail of the method implementation in the plugin \\
\texttt{REQ\_SOFTWARE} & optional & list of external software or libraries necessary for the plugin code \\
\texttt{REQ\_HARDWARE} & optional & describes the resources necessary for the value computation\\
\texttt{ADVICE} & optional & gives an advice on when to use this plugin, instead of other plugins for computing the same set of attributes \\
\midrule
\texttt{compute(entity, **kwargs)} & required & runs the computation for a single entity, given an ID or filename for the entity and optional parameters \\
\texttt{initialize() -> state} & optional & initializes a state variable, which can be passed to the compute function and persists between its calls \\
\texttt{finalize(state)} & optional & performs finalization operations on the persistent state variable \\
\bottomrule
\end{tabular}

\caption{Metadata, interface description and functions required / optionally provided by Prenacs plugins}
\label{tab:plugins}
\end{table*}

\subsubsection{Entity input set}

For each batch computation, the input set of entities must be defined.
Each entity has an unique ID. Often the entity data is stored in files.
The system is flexible so that inter-dependencies of entity IDs and filenames
can be defined using Python functions. Thus the entity set can be e.g. defined
by referring to a list of identifiers in a file or database table, or by
pointing to directory paths containing input files for each entity.

The input of the computation can be defined incrementally, in order to
avoid re-computation for entities, for which a value had already been determined.
The IDs to skip can be provided directly or computed using a Python function.

\subsubsection{Computation parameters}

Computation parameters for a batch computation run can be provided as a YAML file.
The parameters are passed to the \texttt{compute()} function as optional keyword
arguments. The type and meaning of the parameters must be described in the
plugin metadata (\texttt{PARAMETERS} constant).
The computation parameters are assumed to remain constant for each call of the
compute function.

\subsubsection{Persistent state data}

Sometimes common resources are needed by multiple instances of a batch
computation For example, it could be necessary to parse data files, to load
some data into memory, or to initialize a connection to a database. Or some
statistics could be collected during the computation and output at the end.
This data is passed to computation function using a special argument
name \texttt{state}. The state variable can be created by
an initialization function, to which a set of initialization parameters can be provided.
Differently from the computation parameters, the state is allowed to be modified
by the computation function, and persists between calls.
If necessary, a finalization function can be defined,
to perform operations on the state data after the last call of the compute function.
Using a state can be more complex in the case of parallel computations,
where it must be considered how the resources
pointed by the state variable could be concurrently accessed
by parallel instances of the compute function.

\subsubsection{Batch computation results}

A batch computation can be started using the provided command line tools,
or triggered from inside other Python programs, using the Prenacs application program interface. 
Parallel computation is supported, on the local computer, based on the multiprocessing Python module,
or by running the computation on a computer cluster managed by Slurm.

The results of each computation are sets of single or multiple values (scalar or compound)
for different attributes of the input entities.
For performance reasons, the computation results are not stored immediately in the
database. They are instead batch loaded into the database after the computation, alongside
the provenance tracking information.

For each batch computation, an UUID is generated, which is stored alongside each results unit.
A table stores the metadata of the batch computations, including input size, running time measurements,
user and system identifiers, and references to the plugin used for the computation (name and version).
The plugins database table stores the metadata for each plugin in each version.
Automatic computation pipelines are supported in that batch computations metadata also include
a reason variable, which describes why the computation was run (availability of new entities,
availability of new attributes, or recomputation after update of the computation method).

\begin{figure*}
\ \vspace{5mm}
\hrule
\ \vspace{0.5mm}
\begin{verbatim}
# Python
def compute(entity, **kwargs):
  ...

# Nim
proc compute(entity: string, p1: t1 = d1, p2: t2 = d2):
             tuple[results: seq[string]], logs: seq[string]] {.exportpy.}

// Rust
#[pyfunction] fn compute(filename: &str) -> PyResult<(...)> { ... Ok(...)

# Bash
function compute() { local filename=$1; shift; kwargs=$* ...
\end{verbatim}
\ \vspace{-1mm}
\hrule
\ \vspace{-2mm}
\caption{Signature or beginning of compute functions implemented in Python, Nim,
Rust and Bash plugins. In Nim all accepted parameters 
must be listed (in the example, two parameters are defined). In Bash, parameters
are passed as strings of the form \texttt{key=value}, whose contents are evaluated.}
\label{fig:computefn}
\end{figure*}

\subsection{The AttrTables attribute values storage system}

AttrTables is a library which manages a database storing system for any number of
attributes of a collection of entities. Thereby, entities are rows in database tables,
identified by a unique primary key. Each entity has one or multiple attributes,
which are stored in database columns.

AttrTables is implemented in Python. It is based on the SqlAlchemy library \citep{sqlalchemy},
to interface with the database, which is managed by MariaDB \citep{mariadb}.

The main class of the library (\texttt{AttributeValueTables} 
represents the collection of tables, where the attributes are stored.
It allows for dynamic definition of attributes of different datatypes by
creating a set of multiple tables under the hood,
designed to efficiently store and retrieve the computation data and
metadata.
When using the class, the user does not need to know in which of the table a given
attribute is actually stored (Figure \ref{fig:attrtables}).

The AttrTables package is available in Github at
\url{https://github.com/ggonnella/attrtables} and on the PyPi repository
at \url{https://pypi.org/project/attrtables}. This text refers to version 1.2.

\subsubsection{Attribute definition and storage}

The attribute columns are automatically spread among the database tables, so that the total
number of columns does not generally exceed a given limit (by default 64).
New attributes can be defined at any moment, or existing attribute deleted:
The set of tables is automatically
managed, and accessed in a way that is transparent for the user.

Before values can be stored for an attribute, the attribute must be defined.
Each attribute has a name, a datatype and optionally can be assigned to a
computation group. Attributes can be scalars (single values) or compound,
i.e. arrays of values of the same type (e.g. \texttt{Integer[10]}), or tuples of values of different type
(e.g. \texttt{Integer;Float;String(50)}).

Once an attribute has been added, values of the attribute for a number of
entities can be set, and attribute values can be queried,
without knowing in which table the attribute is physically stored.

\subsubsection{Provenance tracking support}

Alongside with the values, each attribute is associated to a computation ID,
which is intended to be a reference to an external table of computation metadata.
Besides for individual attributes, groups of attributes can be defined:

If for an entity, all attributes of a group are computed at once, the computation ID
of the group is stored instead of the single IDs, saving place (since all the
computation IDs of the attributes will be set to NULL)

\subsubsection{Batch computation support}

For performance reasons, the results of a batch computation can be directly
loaded from a tab-separated file, instead of adding them singularly from the
Python code. Thereby a temporary table is created, the data is loaded to the table
and then merged with the original tables.

\begin{figure*}
\ \vspace{5mm}
\hrule
\ \vspace{0.5mm}
\begin{verbatim}
engine = create_engine(connection_string, future=True)
connection = engine.connect()
avt = AttributeValueTables(connection)
avt.create_attribute("a1", "Integer[2]")
avt.create_attribute("a2", "Float[2]")
avt.set_attributes(["a1", "a2"], {"e1": [1, 2, 1.1, 2.2]})
avt.unset_attribute("a1", ["e1"])
avt.load_computation('comp1', ["a2", "a3"], "comp1.tsv")
avt.query_attribute("a3", ["e1","e2","e3"])
avt.destroy_attribute("a3")
\end{verbatim}
\ \vspace{-1mm}
\hrule
\ \vspace{-2mm}
\caption{Example usage of the AttrTables library. The user does not need to known in which of the underlying database table
the data is actually stored.}
\label{fig:attrtables}
\end{figure*}

\subsection{The ProSt package}

The ProSt package is the main package of the ProSt system.
It provides scripts and pipelines for the computation and storage of attributes
of prokaryotic genomes, making use of the packages presented in the
previous sections (MultiPlug, Prenacs, AttrTables).

The core of the ProSt system is a database, named ProStDB,
managed using the MariaDB database system \citep{mariadb}.
The database can be created, filled with data and kept up-to-date using
the pipeline scripts provided with the ProSt package.

ProStDB collects and stores metadata about genome assemblies and group of organisms,
such as taxonomy and phenotype data. Furthermore, using AttrTables,
for each genomes, attributes can be stored, i.e. any kind of information
collected or computed about the genomes. Alongside the attribute values,
provenance tracking information about the attribute computation is stored.

The ProSt package is available in Github at
\url{https://github.com/ggonnella/prost}. This text refers to version 1.0.

\subsubsection{ProSt package components}

The ProSt package is organized in different components,
which allow to create, update, compute new attributes, setup and interact with the database
and analyse the data contained in it (Figure \ref{fig:prostarch}).

The highest level component of ProSt is a collection of pipeline scripts, called \textit{ProStSnakes}.
It is implemented using SnakeMake \citep{Snakemake}. The pipeline scripts execute a number of automated tasks:
automatically download and keeping
the data and metadata obtained from external databases up-to-date; setting up the ProStDB database
and storing the downloaded
data in it; performing computations of any number of attributes using Prenacs, storing the results in ProStDB;
performing comparison tasks based on scripts in the ProStPy library.
User configuration files can be stored in different positions in the file system, according to the documentation;
the automatic retrieval of the configuration was abstracted into a separate package called \textit{findconfig}.

The ProStPy library is a Python package which includes scripts and modules, implementing the
operations executed by the pipeline scripts. The library supports the operations on the ProStDB database and on genomic data.
For example some scripts allow
analysing the attribute values distributions for given groups of organisms.
Each script has a command line interface, as well as a API for SnakeMake: This dual
interface was abstracted into a separated package called \textit{SnaCLI}.

Attributes of genomes are computed using the Prenacs library. ProSt contains a collection
of MultiPlug plugins \textit{ProStPlugins}, written in Python, Nim and Rust, which can be used in Prenacs.
They allow computing attributes or collecting existing values from database
about the assemblies stored in the ProStDB. Only some examples are provided:
The user can write own plugins, according to the provided documentation.

\section{Discussion}

In this manuscript, we presented a system for the definition, computation and storage of genome attributes,
named ProSt (short for ``Prokaryotic Statistics''). The system is implemented in Python, as a series
of different packages. Here we discuss the most important goals that we achieved in the engineering of the system.

\subsection{Dynamic attribute computation system}

ProSt was designed in order to support the verification and discovery of rules which describe the
expected genome contents in group of prokaryotic organisms. In our framework, introduced
in the manuscript \citet{unambiguous}, genome contents are defined in terms of attributes, i.e.
single quantitative aspects of the genome annotation or sequence.

Attributes of genomes can be defined in many different ways.
For example, in \citet{collection} we presented
a collection of annotation rules, for which hundreds of attributes were defined.
Many more thousands attributes could be defined, e.g. as the presence or count
of single genes or groups of genes. Thus, one engineering goal in the development of ProSt
was to have an open system, in which new attributes could be defined in any moment.

Computing the value of genome attributes is a very heterogeneous task. In some cases,
the value can be retrieved from an external source. In this case a simple Python
or Bash script can be written for the task. In other cases, a more complex computation
is necessary. In such cases a program could be written in compiled languages,
such as Nim or Rust, which allow achieving a better performance.
The necessary flexibility for the attribute definition and computation was obtained by implementing our
multi-language plugin system, the library MultiPlug. The system was integrated in our batch computation library, Prenacs, so that
whenever new attributes are defined, plugins can be written for their computation.

Once the attribute values are computed, they must be stored.  Due to the plugin system,
the list of the attributes and their data types are not known in advance.
An existing solution for storing dynamic attributes like these is an entity-value model.
In this, there is a single table, in which there are columns for the entity ID,
the attribute name and the attribute value. Since the values for different attributes
are all stored in the same column, they must be stored using a generic data type (e.g. binary
blobs) and the application must convert back and forth to the correct type.
Furthermore, indexing is generally not possible. Similar performance penalties apply to NoSQL
storage solutions, for which entities can have any number of attributes.
Thus instead, we created an efficient, yet dynamic, attribute storage system, called
AttrTables. This creates new columns for each attribute. To avoid an excessive
number of columns, the attributes are automatically spread across multiple
tables, which are automatically managed.

\subsection{Batch computation and attribute values storage with provenance tracking}

Reproducible research is a key feature, which must be ensured in all types of scientific investigations.
This is sometimes challenging in fields, such as applied bioinformatics and genomics, where
results are often based on heuristics, which differ depending on the input data but also 
on the exact set of software version, parameters and methods.

Several solutions are routinely used for documenting the production of results.
E.g.\ interactive computations can be documented and often reproduced by storing
the computation history in Jupyter notebooks \citep{jupyter}. This solution is ideal,
when a fixed set of computations is run, e.g. for analysing a given dataset. In
contrast, it is not well suited for a dynamic input, such as entries in a
biological sequence database, for which new computations are incrementally run
from time to time.

Pipeline management systems, such as Snakemake \citep{Snakemake} can be used
to define workflows for the analysis, which can be re-applied easily to new
data. However, such systems do not usually define a system for storing the results,
alongside with the parameters and metadata of the code used for computing them.

When developing Prenacs, we took particular care to offer a solution to the
provenance tracking problem. Thus, the package provides a flexible and
extensible system for batch computations, allowing for incremental computations
and changes in the computation code, combined with a database storage solution
which records the provenance of each result.

\begin{acknowledgements}
Giorgio Gonnella has been supported by the DFG Grant GO 3192/1-1 ‘`Automated characterization of microbial genomes and metagenomes by collection and verification of association rules’’. The funders had no role in study design, data collection and analysis, decision to publish, or preparation of the manuscript.
\end{acknowledgements}

\begin{contributions}
 These contributions follow the Contributor Roles Taxonomy guidelines: \href{https://casrai.org/credit/}{https://casrai.org/credit/}.
 Conceptualization: G.G.;
 Data curation: G.G.;
 Formal analysis:  G.G.;
 Funding acquisition:  G.G.;
 Investigation: G.G.;
 Methodology: G.G.;
 Project administration: G.G.;
 Resources: G.G.;
 Software: G.G.;
 Supervision: G.G.;
 Validation: G.G.;
 Visualization:  G.G.;
 Writing – original draft: G.G.;
 Writing – review \& editing: G.G.
\end{contributions}

\begin{interests}
 The authors declare no competing financial interests.
\end{interests}

\bibliography{references}

\end{multicols}

\end{document}